\documentclass[12pt,showpacs,amssymb,superscriptaddress]{revtex4}
\usepackage{mathrsfs}

\usepackage{amssymb,amsmath}
\usepackage{amsfonts}

\def\d{{\rm d}}

\begin{document}

%%%%%%%%%%%%%%%%%%%%%%%%%%%%%%%%%%%%%%%%%%%%%%%%%%
%%%%%%%%%%%%%  TITLE   %%%%%%%%%%%%%%%%%%%%%%%%%%%

\title{{Fock} quantization of a
scalar field with time dependent mass on the
three-sphere: unitarity and uniqueness}
\author{Jer\'onimo Cortez}\email{jacq@fciencias.unam.mx}
\affiliation{Departamento de F\'\i sica,
Facultad de Ciencias, Universidad Nacional Aut\'onoma de M\'exico,
M\'exico D.F. 04510, Mexico.}
\author{Guillermo A. Mena Marug\'an}\email{mena@iem.cfmac.csic.es}
\affiliation{Instituto de Estructura de la Materia,
CSIC, Serrano 121, 28006 Madrid, Spain.}
\author{Jos\'e M. Velhinho}\email{jvelhi@ubi.pt}
\affiliation{Departamento de F\'{\i}sica, Universidade
da Beira Interior, R. Marqu\^es D'\'Avila e Bolama,
6201-001 Covilh\~a, Portugal.}

\begin{abstract}
We study the Fock description of a quantum free field
on the three-sphere with a mass that depends
explicitly on time, also interpretable as an
explicitly time dependent quadratic potential.  We
show that, under quite mild restrictions on the time
dependence of the mass, the specific Fock
representation of the canonical commutation relations
which is naturally associated with a massless free
field provides a unitary dynamics even when the time
varying mass is present. Moreover, we demonstrate that
this Fock representation is the only acceptable one,
up to unitary equivalence, if the vacuum has to be
$SO(4)$-invariant (i.e., invariant under the
symmetries of the field equation) and the dynamics is
required to be unitary. In particular, the analysis
and uniqueness of the quantization can be applied to
the treatment of cosmological perturbations around
Friedmann-Robertson-Walker spacetimes with the spatial
topology of the three-sphere, like e.g. for
gravitational waves (tensor perturbations). In
addition, we analyze the extension of our results to
free fields with a time dependent mass defined on
other compact spatial manifolds. We prove the
uniqueness of the Fock representation in the case of a
two-sphere as well, and discuss the case of a
three-torus. \vskip 3mm \noindent
\end{abstract}

\maketitle
\newpage
\renewcommand{\thefootnote}{\fnsymbol{footnote}}

\section{Introduction}
\label{int}

The quantization of a classical system is not a
uniquely determined procedure. There exist ambiguities
in different steps of the process which generally lead
to inequivalent quantum theories and, hence, to
different quantum physical descriptions. In standard
quantum mechanics, for systems with a finite number of
degrees of freedom, these ambiguities are usually
resolved by adopting a strongly continuous,
irreducible, and unitary representation of the Weyl
algebra associated with the linear phase space. The
Stone-von Neumann theorem \cite{Vonneu} guarantees the
uniqueness of the representation up to unitary
equivalence, which {does} not affect the physical
content of the so-constructed quantum theory. However,
for infinite dimensional systems, an analog theorem
does not exist in generic situations. For instance,
for linear quantum fields and {assuming that it is
possible to adopt} a Fock quantization, where a
concept of particle is available, it is well known
that there exists an infinite number of inequivalent
Fock representations of the Weyl relations
\cite{wald}. In the case of Klein-Gordon fields in
Minkowski spacetime, the symmetry of the background
spacetime is so large that it provides a criterion to
select a unique representation, modulo unitary
equivalence. Namely, one adopts the unique Fock
representation based on a Poincar\'e invariant
physical state, which can viewed as the vacuum.
Poincar\'e invariance suffices to select an
essentially unique complex structure, which is the
mathematical structure which encodes the physical
freedom existing in the Fock quantization (see e.g.
Ref. \cite{schro-fock} for a discussion). For less
symmetric background spacetimes, one cannot appeal to
Poincar\'e invariance in order to arrive at a unique
Fock representation. Even so, provided that the
spacetime is at least stationary, one can still select
a preferred complex structure for a Klein-Gordon field
by imposing some suitable energy criterion
\cite{ash-mag}.

For generic curved spacetimes or for linear scalar
fields with a time dependent mass (or, equivalently,
subject to a time dependent quadratic potential), the
stationarity is lost. Then, in general, no uniqueness
criterion exists and many unitarily inequivalent Fock
representations are possible, so that the quantum
field physics is not uniquely determined. In spite of
this, some uniqueness theorems have been obtained
recently for linear fields in the context of quantum
cosmology \cite{ccmv,ccmv2,cmsv,cmvS2} invoking not
only the symmetries of the field equation, which are
severely restricted for general curved spacetimes, but
also the unitarity of the quantum field dynamics. In
more detail, the uniqueness of the Fock representation
has been proved for a linear free field with time
dependent mass defined on the circle
\cite{ccmv,ccmv2}, or on the two-sphere under the
requirement of axisymmetry \cite{cmsv,cmvS2}.

Apart from the academical interest of these examples
of uniqueness theorems, their physical interest comes
from the symmetry reduction of general relativity by
two spacelike Killing vectors for compact spatial
topologies. This symmetry reduction leads to a family
of cosmological models with inhomogeneities which was
first studied by Gowdy \cite{gowdy}. Gowdy classified
the possible compact topologies, demonstrating that
the spatial sections of the spacetime must be
homeomorphic to either the three-torus $T^3$, the
three-sphere $S^3$, or the three-handle $S^2\times
S^1$. The Gowdy cosmologies can be interpreted as a
compact homogeneous Bianchi universe containing
gravitational waves. In the case of linearly polarized
waves, the infinite number of degrees of freedom
present in the inhomogeneities can be described in
terms of a scalar field which satisfies a wave
equation with a time dependent mass term \cite{cmsv}.
For Gowdy cosmologies with the topology of a
three-torus, the wave equation is that corresponding
to a static $(1+1)$-dimensional spacetime whose
spatial manifold is a circle (equipped with the
standard metric) {\cite{ccmv,ccm}}. For the rest of
possible topologies in the Gowdy models, the wave
equation corresponds to a static $(1+2)$-dimensional
spacetime, the spatial manifold being a two-sphere
\cite{cmvS2,BVV2}. In this later case, in addition,
the field which describes the gravitational waves has
to be axisymmetric.

The Fock representation selected by the criteria of
invariance of the vacuum under the symmetries of the
field equation and unitarity of the dynamics is the
one which is naturally associated with a free massless
scalar field. In the proof of these uniqueness
theorems, a fundamental role is played by the
compactness of the spatial sections of the spacetime
in which the field propagates. A natural question is
whether these uniqueness results can be extended to
linear fields with time dependent mass propagating in
more realistic spacetimes, maintaining the assumption
of compact spatial topology. In particular, one is
interested in $(1+3)$-dimensional spacetimes,
corresponding to the actually observed dimensionality.
This would allow the application of the results to
physically relevant scenarios without the recourse to
a lower dimensionality motivated from symmetry
reductions of general relativity.

In this work, we will focus our discussion on a linear
field with time dependent mass defined on $S^3$. The
interest of this case is manifold. Obviously, the
analysis has a straightforward application for the
quantization of test fields on a
Friedmann-Robertson-Walker (FRW) background with the
spatial topology of a three-sphere. For a Klein-Gordon
scalar field, e.g., a rescaling by the scale factor of
the FRW universe transforms the field equation to the
form of a linear wave equation with time dependent
mass in a static $(1+3)$-spacetime whose spatial
manifold is $S^3$ (with the standard metric). Besides,
in a suitable Lorentz gauge, the sourceless Maxwell
equations on a FRW background with $S^3$ spatial
topology lead to field equations for the vector
potential that are of the considered form (i.e., wave
equations for the propagation in the sphere) with a
vanishing mass \cite{jantzen}.

But the framework in which the discussion finds
probably the most natural application is in the
treatment of perturbations of FRW spacetimes. The
study of perturbations around classical FRW
cosmologies at first order (i.e., keeping in the field
equations only linear terms in the perturbations) was
originally developed by Lifshitz \cite{lif}. An
attempt to provide a description of the perturbations
free of the physical ambiguities introduced by gauge
freedom was made by Hawking \cite{haw}. Later, Olson
obtained a covariant formulation of the perturbative
equations for the case of an isentropic perfect fluid
in a background with vanishing spatial curvature
\cite{olson}. A truly gauge invariant formalism for
the analysis of generic cosmological perturbations was
constructed by Bardeen \cite{bar,bar2}. Without trying
to be exhaustive in the literature about cosmological
perturbations, other aspects of linear perturbations
of FRW spacetimes can be found in Refs.
\cite{muk2,muk}, specially for the case of a scalar
field with arbitrary potential in a flat FRW
background.

In particular, for isotropic perturbations of the
energy-momentum tensor (like e.g. for a perfect fluid)
which are besides adiabatic, the gauge invariant
energy density perturbation amplitude, or equivalently
the (only independent) Bardeen potential, satisfies a
second order linear differential equation which, after
a rescaling by a suitable time function, can be
written as a wave equation with a time dependent mass
term \cite{bar,muk2}. This wave equation is that of a
static and homogenous $(1+3)$-spacetime with the same
spatial topology as the FRW background. In the case of
a FRW universe whose spatial sections are
three-spheres, one arrives to a wave equation for a
field on $S^3$, precisely the case that we will
consider in detail in this work. Furthermore, the
tensor perturbations of a FRW spacetime, describing
the content of gravitational waves, turn out to
satisfy as well a field equation of the considered
form if the perturbations of the energy-momentum
tensor are isotropic: after a rescaling of their
amplitudes by the FRW scale factor (and adopting
conformal time), one arrives at a wave equation for
the tensor perturbations which includes a time
dependent mass \cite{bar}. Again, when the spatial
topology of the FRW universe is a three-sphere, these
perturbations are given by a tensor field on $S^3$.

For the case of a scalar field with nonvanishing
(constant) mass as the matter content, a detailed
study of the perturbations of a FRW spacetime with the
spatial topology of $S^3$ was carried out by Halliwell
and Hawking in the context of quantum cosmology
\cite{hh}. From that work {one can obtain} (apart
from the results about tensor perturbations commented
above) the equation satisfied by the perturbations of
the matter field {expanded} in harmonics on the
three-sphere \cite{jantzen}, {in a gauge where the
perturbations of the spatial metric contain no tensor
harmonics derived from the scalar ones \cite{hh}.
After} rescaling the matter field perturbations by the
FRW scale factor, a careful analysis of the resulting
equation in conformal time shows that the solutions
for large harmonic number $n$ (namely, for large
{negative} eigenvalue of the Laplace-Beltrami
operator on $S^3$) reproduce in fact those for a free
field on $S^3$ in the presence of a time dependent
mass term, up to asymptotic corrections which do not
affect the discussion presented in the rest of this
work.

Then, the uniqueness that we will prove for the Fock
representation of a linear scalar field with time
varying mass implies that, in all the examples
commented above (and after adopting the rescaling of
the field indicated to arrive at the studied
equation), one can select a preferred Fock
quantization, picked up by the symmetries of the field
equation and the unitary implementation of the
dynamics. Furthermore, we will discuss possible
extensions of our uniqueness result for compact
spatial manifolds other than the three-sphere. In
particular, we will demonstrate that the uniqueness is
reached as well for a field on the two-sphere, without
any need to impose axisymmetry (therefore generalizing
the results of Ref. \cite{cmsv}). In addition, we will
comment on the case of a three-torus, which finds
application in the treatment of perturbations of a FRW
spacetime with a flat and compact spatial topology,
according to our comments above.

The rest of the paper is organized as follows. Sec. 2
contains some basic considerations about fields on
$S^3$ and their quantization with Fock techniques. We
then show in Sec. 3 that the Fock representation
associated with a free massless field provides a
unitary implementation of the dynamics even if a time
dependent mass term is present in the field equation
(under quite mild conditions on the time dependence of
the mass). Sec. 4 is devoted to prove that, if one
restricts oneself to complex structures with the
invariance of the field equation and requires a
unitary dynamics, all possible Fock representations
are unitarily equivalent. In this sense, the
quantization is unique. The generalization of the
unitarity and uniqueness results for fields defined on
compact spatial manifolds other than $S^3$ is
discussed in Sec. 5, where we also present the
conclusions. Finally, an appendix is added which
contains some calculations needed in the proof of
uniqueness.

\section{Quantum scalar field on $S^3$}
\label{ccmv-quant}
\subsection{The field and its decomposition in harmonics}
\label{secfi}

{Let us consider a scalar field $\phi$} propagating
in a globally hyperbolic, static $(1+3)$-background
whose spatial manifold is the three-sphere $S^3$
equipped with the standard metric, i.e., the metric
induced from the Euclidean metric on $E^4$:
\begin{equation}
\label{three-sphere-metric} h_{ab}=\d \chi_{a}\otimes
\d\chi_{b}+\sin^{2}\left(\chi\right)\d
\theta_{a}\otimes \d \theta_{b}+
\sin^{2}\left(\chi\right)\sin^{2}\left(\theta\right)
\d \sigma_{a} \otimes \d \sigma_{b}.
\end{equation}
Here, $\sigma\in S^{1}$ and the rest of coordinates
have a range of $\pi$. The $(1+3)$-dimensional
spacetime has the topology of $\mathbb{I} \times S^3$,
where $\mathbb{I}$ is an interval of the real line,
and its metric is $g_{ab}=- \d t_a \d t_b+h_{ab}$.

The scalar field satisfies a linear wave equation
which includes a time dependent mass term, namely, a
potential of the form {$V(\phi)=f(t)\phi^{2}/2$:}
\begin{equation}
\label{sn-fieldequation}
\ddot{\phi}-\Delta\phi + f(t)\phi =0,
\end{equation}
where $\Delta$ denotes the Laplace-Beltrami operator
on $S^{3}$, {$f(t)$ is a real function on
$\mathbb{I}$,} and the dot stands for the time
derivative.

The canonical phase space $\Gamma$ is the space of
Cauchy data at some fixed time $t_0$,
$\{(\varphi,P)\}=\{(\phi_{|t_0},\sqrt{h}
{\dot\phi}_{|t_0})\}$ where
$h=\sin^{2}(\theta)\sin^{4}(\chi)$ is the determinant
of the metric on the spatial section $S^{3}$. The
symplectic structure (independent of the choice of
time section) is
\begin{equation}
\Omega[(\varphi_{1},P_{1}),(\varphi_{2},P_{2})] =\oint
d ^{3} x \, \left(\varphi_{2}P_{1}
-\varphi_{1}P_{2}\right).\end{equation} The
corresponding nonzero Poisson brackets are
$\{\varphi(x),P(x')\}=\delta(x-x')$, where $\delta(x)$
is the Dirac delta on $S^3$, and $x$ collectively
denotes the (hyper)spherical coordinates on $S^3$,
$x:=(\chi,\theta,\sigma)$.

It is clear from the field equation
(\ref{sn-fieldequation}) that the group of rotations
$SO(4)$ is a group of symmetries of the field
dynamics, since the metric, and therefore the
Laplacian, is $SO(4)$-invariant. Upon quantization, we
will therefore look for a unitary implementation not
just of  the field dynamics, but of the group of
symmetries $SO(4)$ as well.

The most natural way to obtain a unitary
implementation of a group of symmetries, or more
generally of any set of symplectic transformations, is
to define the quantum representation of the canonical
commutation relations (CCR's) by means of a state of
the Weyl algebra (interpretable as a vacuum) which is
invariant under the transformations in question. In
the case of $SO(4)$, a representation which is well
known to be invariant is the massless free field
representation, since it is defined by a complex
structure in phase space which is determined
exclusively by the metric and the Laplacian.

As we will see, the dynamics of our field turns out to
be implemented also unitarily in the free field
representation for any (sufficiently regular) function
$f(t)$ in  the field equation
(\ref{sn-fieldequation}).

In order to discuss these issues in more detail, it is
convenient to adopt a description of the field in
terms of harmonics, i.e., solutions of the eigenvalue
equation for the Laplace-Beltrami operator
\cite{harri}. As {it} is well known, fields on
$S^3$ admit a decomposition in (hyper)spherical
harmonics \cite{jantzen,lif,GS,hh}, just like in the
more familiar $S^2$ case. Thus, the scalar field
$\phi(t,x)$ can be written as a linear combination of
modes $Q_{n \ell m}(x)$, which are eigenfunctions of
the Laplace-Beltrami operator $\Delta$ with eigenvalue
$-n(n+2)$, where $n=0,1,2,\ldots$ Explicitly,
\begin{equation}
\label{deco}
\phi(t,x)=\sum_{n, \ell , m }A_{n \ell m }(t)Q_{n \ell m }(x),
\end{equation}
where the coefficients $A_{n \ell m}$ are functions of
time and $\ell$ and $m$ take the values $\ell
=0,\ldots,n$ and $m=-\ell,\ldots,\ell$, respectively.
The scalar harmonics $Q_{n \ell m }$, normalized with
respect to the volume element on $S^3$, have the form
\begin{equation}
\label{sh} Q_{n \ell m
}(\chi,\theta,\sigma)=2^{\ell+\frac{1}{2}}(\ell !)
\sqrt{
\frac{(n-\ell)!(n+1)}{\pi(n+\ell+1)!}}\sin^{\ell}(\chi)\,
C_{n-\ell}^{(\ell +1)}[\cos (\chi)]Y_{\ell
m}(\theta,\sigma),
\end{equation}
where $Y_{\ell m}$ are the usual spherical harmonics
on $S^{2}$ and $C_{n-\ell}^{(\ell +1)}[\cos (\chi)]$
{denote} the Gegenbauer polynomials \cite{abra,GR}.
With fixed $n$, the functions $Q_{n\ell m}$ span an
irreducible representation of $SO(4)$ of dimension
$(n+1)^2$ (see e.g. Ref. \cite{jantzen}).

Let us restrict our discussion to the  particular case
of a real field. We start by {writing} down an
alternative expression for the decomposition
(\ref{deco}) {of a (generally complex) field $\phi$
which is better adapted to the real case,}
\begin{equation}
\label{real} \phi=\sum_{n,\ell}q_{n\ell 0}\, Q_{n\ell
0}+ \sqrt{2}\sum_{n,\ell,m>0}q_{n\ell m}\,
\Re[Q_{n\ell m}]+ \sqrt{2}\sum_{n,\ell,m>0}q_{n\ell
-m}\, \Im[Q_{n\ell m}]. \end{equation} Here, the
integer $m$ is strictly positive in the last two sums,
and the symbols $\Re$ and $\Im$ denote, respectively,
the real and imaginary parts. We note first that the
well known behavior of the spherical harmonics under
complex conjugation,
\begin{equation}Y_{\ell m}^*=(-1)^m Y_{\ell -m},
\end{equation}
implies a similar behavior for the scalar harmonics on
$S^3$,
\begin{equation}
\label{conj}
Q_{n\ell m}^*=(-1)^m Q_{n\ell -m},
\end{equation}
because the remaining factors in $Q_{n \ell m }$ are
real and independent of $m$ [see Eq. (\ref{sh})].
{In addition, the functions $Q_{n\ell 0}$ are real
as well. Therefore a general real field is obtained
just by restricting all considerations to real
coefficients in the decomposition (\ref{real}).}

{Thus}, the configuration space of a real scalar
field in $S^3$ is in one-to-one correspondence with
the space of all real coefficients
\begin{equation}\{( q_{n\ell m});\, n=0,1,\ldots;\,
\ell=0,\ldots,n;\,
m=-\ell,\ldots,\ell\}.\end{equation}

Taking into account the field equation
(\ref{sn-fieldequation}), the orthogonality properties
of the harmonics $Q_{n\ell m}$, and relation
(\ref{conj}), it is straightforward to see that these
modes obey completely decoupled equations of motion,
which moreover depend only on $n$. In particular, all
the modes $q_{n\ell m}$  with the same $n$ satisfy the
same equation of motion:
\begin{equation} \label{q-eq} \ddot
q_{n\ell m}+\left(\omega_n^2+f\right)q_{n\ell m}=0,
\end{equation}
where
\begin{equation}
\label{omega} \omega_n^2:=n(n+2).
\end{equation}
Note that, if we call $k:=n+1$, we have
$\omega^2_k=k^2-1$ with the above definition.
Actually, we might absorb the contribution of $-1$ in
$\omega^2_k$ by redefining $f(t)$, and then work in
the unit mass representation, with $\omega_k=k$. Since
this will obscure the interpretation of the quantities
appearing in our discussion, we will continue to use
the mode number $n$ as our label and the original
function $f$ in our equations.

So, as we have seen, the modes $q_{n\ell m}$ describe
completely decoupled degrees of freedom. We will
{call} ${\cal Q}_n$ the corresponding subset of the
full configuration space associated with a fixed $n$,
i.e., ${\cal Q}_n$ is the linear space of dimension
$(n+1)^2$ spanned by the modes $q_{n\ell m}$ with the
same label $n$ (while $\ell\in\{0,\ldots, n\}$ and in
turn $m\in\{-\ell,\ldots,\ell\}$).

The momentum variables canonically conjugate to the
above configuration variables are $p_{n \ell
m}=\dot{q}_{n \ell m}$. From the basic Poisson bracket
between $\varphi$ and $P$ and the orthogonality of the
scalar harmonics, one can check that $q_{n \ell m}$
and the introduced $p_{n \ell m}$ form a complete set
of variables in phase space which is canonical, i.e.,
\begin{equation}\{q_{n\ell
m},p_{n'\ell'm'}\}=\delta_{nn'}\delta_{\ell\ell'}
\delta_{mm'},\end{equation} with all other Poisson
brackets between this set of variables being equal to
zero.

Clearly, each space ${\cal Q}_n$, as well as the
corresponding momentum space ${\cal P}_n$, carries an
irreducible representation of $SO(4)$ of dimension
$(n+1)^2$. Moreover, the representations realized in
${\cal Q}_n$ and in ${\cal P}_n$ are actually the
same.$^{\footnotemark[1]}$\footnotetext[1]{{This
means that the matrices representing $SO(4)$-rotations
are the same in ${\cal Q}_n$ and ${\cal P}_n$, if one
adopts the respective bases provided by $\{q_{n\ell
m}\}$ and $\{p_{n\ell m}\}$.}}

\subsection{Fock Quantization}
\label{repsec}

One can put forward a Schr\"{o}dinger representation
of the CCR's on a Hilbert space ${\cal H}=L^2({\cal
Q},\mu)$ of square integrable (complex-valued)
functions on the infinite dimensional linear space
${\cal Q}$ given by the direct sum of all the spaces
${\cal Q}_n$. For the sake of simplicity in the
presentation, we will drop from now on the $n=0$
mode.$^{\footnotemark[2]}$\footnotetext[2]{It is a
single mode, decoupled from the remaining degrees of
freedom, {and which can be quantized, at least for
\emph{nonnegative} mass functions $f(t)$,} using the
standard Schr\"{o}dinger representation, with the
Lebesgue measure on $\mathbb{R}$. {Besides}, the
action of $SO(4)$ on this mode is trivial. Therefore,
{provided that $f(t)$ is nonnegative,} the $n=0$
mode plays no role in {the} discussion of unitarity
and uniqueness of the quantum representation. The full
Hilbert space for the scalar field is {then} the
tensor product of $L^2(\mathbb{R},dq_{000})$ with
$\cal H$.} The measure $\mu$ is the Gaussian measure
on $\cal Q$, obtained from the product of the
1-dimensional Gaussian measures, one per each degree
of freedom, defined by the frequencies $\omega_n$.
More precisely, the measure is
\begin{equation}
\label{mu} \d \mu=\prod_{n, \ell,
m}\left(\sqrt{\frac{\omega_n}{\pi}} \,
e^{-\omega_nq_{n\ell m}^2} \, \d q_{n\ell m} \right).
\end{equation}
The ($\mu$-compatible) basic operators of
configuration and momentum act as multiplicative and
derivative operators, respectively,
\begin{equation}
\label{rep} \hat q_{n\ell m} \Psi = q_{n\ell m}\Psi
,\qquad\qquad \hat p_{n\ell m}\Psi = -i
{\frac{\partial}{\partial q_{n\ell m}}}\Psi +i
\omega_n q_{n\ell m}\Psi,
\end{equation}
where $\Psi\in {\cal H}$ is  arbitrary.

The constructed representation is just the one which
is naturally associated with the free massless field.
In this respect, let us point out that the above
representation is defined by a complex structure $j_0$
on the canonical phase space which takes the {form}
\cite{schro-fock}
\begin{eqnarray}
\label{cano-cs} j_0\left( \begin{array}{c} \varphi \\
P\end{array}\right) = \left( \begin{array}{cc} 0 &
-(-h \Delta)^{-1/2}
\\(-h \Delta)^{1/2} &  0\end{array}\right) \left(
\begin{array}{c} \varphi \\ P\end{array}\right).
\end{eqnarray}

Let us remind that a complex structure is the
mathematical structure that encodes the ambiguity
which is physically relevant in the Fock quantization
(or, equivalently, in its Schr\"{o}dinger
counterpart). A complex structure $J$ is a symplectic
transformation in phase space that is compatible with
the symplectic structure [in the sense that the
combination $\Omega(J\cdot,\cdot)$ provides a positive
definite bilinear map], and such that its square
equals minus the identity, $J^2=-1$ (see Refs.
\cite{wald,schro-fock,otros} for details on the way
{in which} a complex structure determines a Fock
representation --and/or the corresponding
Schr\"{o}dinger one).

The complex structure (\ref{cano-cs}) is obviously
invariant under the action of $SO(4)$, a fact which
immediately leads to a unitary implementation of the
symmetry group. Of course, the complex structure is
also time-translation invariant, a property which
ensures a unitary implementation of the \emph{free}
field dynamics. Note however that no dynamically
invariant complex structure is expected to exist when
the field equation (\ref{sn-fieldequation}) is
considered, for a generic function $f(t)$. We will
then follow the strategy of considering
$SO(4)$-invariant complex structures (there are plenty
of those, as we will see) which, {rather than
being} invariant under the evolution, allow one {at
least} to obtain a unitary implementation of the field
dynamics.

In order to discuss whether the dynamics is unitary,
it is more convenient to replace the canonical
variables $(q_{n\ell m},p_{n\ell m})$ with the
annihilation-like variables
\begin{equation}
\label{basic-var} a_{n\ell
m}={\frac{1}{\sqrt{2\omega_n}}} \left(\omega_nq_{n\ell
m} +i p_{n\ell m}\right), \qquad n=1,2,\ldots,
\end{equation}
and the creation-like variables provided by their
complex conjugates $a^{*}_{n\ell m}$. We will use this
set of complex variables as coordinates for the
inhomogeneous sector ($n\neq 0$) of the canonical
phase space. Note that the variables (\ref{basic-var})
correspond to the annihilation operators of the
considered quantum representation. Precisely because
of that, the complex structure $j_{0}$ given in Eq.
(\ref{cano-cs}) takes a particularly simple form with
respect to these variables: it is  block diagonal and
completely defined by
\begin{equation}
\label{j0}
j_0(a_{n\ell m})=i a_{n\ell m},\qquad
j_0(a^*_{n\ell m})=-i a^*_{n\ell m},
\quad \forall n, \ell, m.
\end{equation}

As a consequence of the decoupling between degrees of
freedom, the finite transformations generated by the
dynamics are linear symplectic transformations which
can be decomposed in $2\times 2$ blocks, one for each
fixed pair $(a_{n\ell m},a^{*}_{n\ell m})$. Thus, the
classical evolution of the annihilation and
creation-like variables from time $t_0$ to time $t$ is
totally determined by a discrete set of $2\times 2$
matrices ${\cal U}_{n\ell m}(t,t_0)$. Furthermore,
since the dynamical equations (\ref{q-eq}) are
independent of $\ell$ and $m$, the same is true for
the evolution matrices, i.e.,
\begin{eqnarray}
\left( \begin{array}{c} a_ {n\ell m}(t)\\ a_{n\ell m}^*(t)
\end{array}\right)= {\cal U}_{n\ell m}(t,t_0)
\left( \begin{array}{c} a_{n\ell m} (t_0)\\
a_{n\ell m}^* (t_0)\end{array} \right)= {\cal U}_{n}(t,t_0)
\left( \begin{array}{c} a_{n\ell m} (t_0)\\
a_{n\ell m}^* (t_0)\end{array} \right),
\end{eqnarray}
\begin{eqnarray}
\label{bogo-transf} {\cal U}_{n\ell m}(t,t_0):=\left(
\begin{array}{cc} \alpha_{n\ell m}(t,t_0) &
\beta_{n\ell m}(t,t_0) \\ \beta_{n\ell m}^*(t,t_0) &
\alpha_{n\ell m}^*(t,t_0) \end{array} \right)= {\cal
U}_n(t,t_0):=\left( \begin{array}{cc} \alpha_n(t,t_0)
& \beta_n(t,t_0) \\ \beta_n^*(t,t_0) &
\alpha_n^*(t,t_0)
\end{array} \right).
\end{eqnarray}
Here, $\alpha_n(t,t_0)$ and $\beta_n(t,t_0)$ are
Bogoliubov coefficients for the evolution equations
(\ref{q-eq}), which therefore depend on the specific
function $f(t)$. Since ${\cal U}_n(t,t_0)$ provides a
symplectomorphism, one can easily check that, for any
value of $t_0$ and $\forall n$,
$|\alpha_{n}(t,t_0)|^{2}-|\beta_{n}(t,t_0)|^{2}=1$ for
all $t\in \mathbb{I}$.

\section{Unitary dynamics}
\label{sec:uni}

We will now show that the field dynamics is
implemented as a unitary transformation in the quantum
representation defined by $j_0$ even if the field
possesses a nonconstant mass, at least under quite
mild conditions on the time dependence of this mass
and in spite of the loss of time invariance of the
system.

\subsection{Conditions for a unitary implementation}

Let us consider the operators ${\hat a}_ {n\ell m}$,
related to the operators ${\hat q}_ {n\ell m}$ and
${\hat p}_ {n\ell m}$ [introduced in Eq. (\ref{rep})]
by means of the linear relations (\ref{basic-var}). As
mentioned above, these are the annihilation operators
of the $j_0$-representation.

In the Heisenberg picture, time evolution in the
quantum theory is, in principle, given again by the
Bogoliubov transformation (\ref{bogo-transf}), what
means that the operators $(\hat{a}_{n\ell m},
\hat{a}^{\dag}_{n\ell m})$ at time $t_0$ evolve to
operators $(\hat{a}_{n\ell m}(t),
\hat{a}^{\dag}_{n\ell m}(t))$ at time $t$ according to
the same transformation:
\begin{eqnarray}
\label{qtime-evol}
\left( \begin{array}{c} \hat{a}_{n\ell m}(t) \\
\hat{a}^{\dag}_{n\ell m}(t)\end{array}\right) = \left(
\begin{array}{cc} \alpha_n(t,t_0) & \beta_n(t,t_0) \\
\beta_n^*(t,t_0) &  \alpha_n^*(t,t_0) \end{array}
\right) \, \left( \begin{array}{c}  \hat{a}_{n\ell m} \\
\hat{a}^{\dag}_{n\ell m} \end{array}\right).
\end{eqnarray}
The question that we want to address is whether  or
not these transformations as a whole ($\forall n$,
$\ell$, and $m$) can be unitarily implemented, i.e.,
if they correspond to a unitary transformation
${U}(t,t_0)$ in the Hilbert space $\cal H$.

The main advantage of using the variables
(\ref{basic-var}) is that the condition for unitary
implementability can be rephrased in a simple form. In
general, a symplectic transformation $R$ can be
unitarily implemented on a Fock representation,
constructed from a complex structure $J$, if and only
if $R+JRJ$ is an operator of the Hilbert-Schmidt type
on the corresponding 1-particle Hilbert space
\cite{hr,sh}. Equivalently, $R$ is implementable as a
unitary transformation if and only if $J- RJR^{-1}$ is
a Hilbert-Schmidt operator (note that each of these
conditions ensures that the representations defined by
the complex structures $J$ and $RJR^{-1}$ are
unitarily equivalent, something which is essentially
the definition of unitary implementability). In the
case of the family of symplectic transformations
defined by the classical dynamics, specified by the
matrices ${\cal U}_n(t,t_{0})$, the Hilbert-Schmidt
condition for a unitary implementation in the
$j_0$-Fock representation becomes a square summability
condition on the coefficients $\beta_{n\ell m}$,
namely,
\begin{equation}
\label{uny} \sum_{n=1}^{\infty}\sum_{\ell
=0}^{n}\sum_{m=-\ell}^{\ell} |\beta_{n \ell
m}(t,t_0)|^{2}<\infty\quad \forall t \in \mathbb{I},
\end{equation}
given a fixed reference time $t_0$.

In the present case, owing to the independence of the
equations of motion (\ref{q-eq}) with respect to
$\ell$ and $m$, we have
\begin{equation}\sum_{n
=1}^{\infty}\sum_{\ell =0}^{n}\sum_{m=-\ell}^{\ell}
|\beta_{n \ell m}(t,t_0)|^{2}=\sum_{n
=1}^{\infty}g_{n}
|\beta_{n}(t,t_0)|^{2},\end{equation} where the
degeneracy factor $g_{n}=(n+1)^{2}$ counts the number
of degrees of freedom with the same dynamics. Thus,
the unitary implementability condition becomes a
square summability condition for the sequences
$\sqrt{g_{n}}\beta_{n}(t,t_0)$, i.e.,
\begin{equation}
\label{condition}
\sum_{n =1}^{\infty}g_{n}|\beta_{n}(t,t_0)|^{2}< \infty,
\end{equation}
where the coefficients $\beta_n$ are those
corresponding to the differential equations
\begin{equation} \label{eqnotlm} \ddot
q_n+\left(\omega_n^2+f\right)q_n=0.
\end{equation}

\subsection{Asymptotic dynamics and unitarity}
\label{asy}

In order to elucidate whether condition
(\ref{condition}) is fulfilled, we are interested in
investigating the large $n$-limit of the coefficients
$\beta_n$, and therefore the behavior of the solutions
to the differential {equation} (\ref{eqnotlm}) for
large $n$.

Let us start by writing the general solution to
{that equation} of motion in the form
\begin{equation} \label{gen}
q_n(t)=A_n\exp[\omega_n\Theta_n(t)]+
A^{*}_n\exp[\omega_n\Theta^{*}_n(t)].
\end{equation}
For each $n$, $A_n$ is a complex constant related to
the initial conditions and $\Theta_n$ is a particular
(complex) solution of the equation
\begin{equation} \label{car} \omega_n\ddot \Theta_n+
\omega_n^2{\dot \Theta_n}^2+\omega_n^2+f=0,
\end{equation}which is obtained from Eqs.
(\ref{eqnotlm}) and (\ref{gen}).

We fix the arbitrariness in the solution $\Theta_n$ of
the above equation by means of the initial conditions
$\Theta_n(t_0)=0$ and $\dot{\Theta}_n(t_0)=-i$,
$\forall n$. A simple analysis shows that this is
always possible (see Ref. \cite{cmsv} for details).
This choice is motivated by the free massless scalar
field case [i.e., the case with $f(t)=0$], for which
$\dot{\Theta}_n=-i$ is satisfied not only initially,
but at all times.

By working out the relation between the canonical data
$(q_n(t_0), {\dot q}_n(t_0))$ and the complex constant
$A_n$ in Eq. (\ref{gen}), one can easily obtain the
evolution matrices in terms of the canonical
variables. Finally, changing from those variables to
the annihilation and creation-like variables, it is
straightforward to deduce the expression of the
Bogoliubov coefficients $\alpha_{n}(t,t_0)$ and
$\beta_{n}(t,t_0)$ appearing in the evolution matrices
(\ref{bogo-transf}). We get
\begin{eqnarray} \label{alpha}
\alpha_n(t,t_0) & = & \frac{1}{2}e^{\omega_n\Theta_n(t)}
\left[1+i\,\dot
\Theta_n(t) \right] ,\\
\label{beta} \beta_n(t,t_0) & = &
\frac{1}{2}e^{\omega_n\Theta^*_n(t)} \left[1+i\,
{\dot\Theta}^*_n(t) \right] .
\end{eqnarray}

As an aside we note that, in the equations of motion
(\ref{eqnotlm}), the term $\omega_n^2$ dominates over
the $n$-independent mass term $f(t)$ in the limit of
large $n$; we thus expect that the solutions $q_n(t)$
converge to those corresponding to the free massless
case (for the same initial conditions), at least for
sufficiently regular functions $f(t)$.

In order to see that the asymptotic corrections to
this behavior for large $n$ do not affect unitarity,
we only need to analyze the unitary implementability
condition (\ref{condition}), which translates now into
\begin{equation}
\label{newcondition} \sum_{n
=1}^{\infty}g_{n}e^{2\omega_n\Re{[\Theta^*_n(t)}]}
|1+i\,{\dot\Theta}^*_n(t)|^{2}< \infty \quad \forall t
\in \mathbb{I}.
\end{equation}
In the following we will therefore focus  on the
behavior of $\Theta_n$ for large $n$.

It is convenient to write the functions
$\dot{\Theta}_n$ in the form
\begin{equation} \label{newze1}
\dot{\Theta}_n=-i+\frac{W_n}{\omega_n},
\end{equation}
so that the expected asymptotic limit, for large $n$,
is already singled out. From Eq. (\ref{car}), it
follows that the functions $W_n$ satisfy the
first-order differential equations
\begin{equation}
\label{car-w} \dot{W}_{n}=2i\omega_n W_{n}- W^{2}_{n}-f,
\end{equation}
with the initial conditions $W_n(t_0)=0$, deduced from
the above conditions on the functions
$\dot{\Theta}_n$. In addition, recalling that
$\Theta_n(t_0)=0$, we obtain
\begin{equation}
\label{zeca} \Theta_n=-i (t-t_0)+ \int^t_{t_0}
\d\bar{t}
\,\frac{W_n(\bar{t})}{\omega_n}.\end{equation}
Condition (\ref{newcondition}) can then be rewritten
as
\begin{equation}
\label{newnewcondition} \sum_{n =1}^{\infty}g_{n}
e^{2\int^t_{t_0}\Re{[W_n]}}\,
\frac{|W_n(t)|^{2}}{\omega_n^2}< \infty \quad \forall
t \in \mathbb{I}.
\end{equation}

We will now prove that, in the large $n$-limit,  the
desired solutions to Eq. (\ref{car-w}) possess
``ultraviolet modes'' of order $1/n$, such that the
unitarity condition (\ref{newnewcondition}) is
satisfied. The argument goes as  follows. In the limit
of large $n$, the term $W^{2}_{n}$ in Eq.
(\ref{car-w}) is dominated by the term linear in
$W_{n}$, whose coefficient grows with $n$ [see
Eq.~(\ref{omega})]. One can then start by neglecting
the quadratic term $W^{2}_{n}$, show that the
resulting linear equation admits solutions $\bar{W}_n$
of order $1/n$, and check that, in the limit of large
$n$, the contribution of the quadratic term for such
solutions is in fact negligible in the original
differential equation.

Let us then consider the linear differential equation
obtained from Eq. (\ref{car-w}) after removing the
quadratic term:
\begin{equation} \label{w-lin}
\dot{\bar{W}}_{n}=2i\omega_n \bar{W}_{n}-f.
\end{equation} The solution
to Eq. (\ref{w-lin}) satisfying the initial condition
$W_n(t_0)=0$ is
\begin{equation}
\bar{W}_{n}(t)=-\exp(2i\omega_nt)\,\int_{t_0}^t \d
\bar{t}\, f(\bar{t})\exp(-2i\omega_n\bar{t}),
\label{lin-sol}
\end{equation}
and an integration by parts leads to
\begin{equation}
\label{newze2}
\bar{W}_{n}(t)=-\frac{if(t)}{2\omega_n}+\frac{if(t_0)\,
e^{2i\omega_n(t-t_0)}}{2\omega_n} -
\frac{\exp(2i\omega_nt)}{2i\omega_n}\,\int_{t_0}^t \d
\bar{t}\, \dot{f}(\bar{t})\exp(-2i\omega_n\bar{t}).
\end{equation}

The absolute value of the last term in Eq.
(\ref{newze2}) is bounded by $\int_{t_0}^t \d t\,
|\dot{f}|/(2\omega_n)$. One therefore concludes that
there is a function $C(t)$, independent of $n$, such
that the absolute value of the solutions
(\ref{lin-sol}) is bounded by $C(t)/\omega_n$. To
reach this conclusion, it is {sufficient} that the
derivative of the function $f(t)$ exists and is
integrable in every closed interval $[t_0,t]$ (or
$[t,t_0]$) of $\mathbb{I}$.

We now return to the original differential equation
(\ref{car-w}). Since the quadratic term
$\bar{W}^{2}_{n}$ is bounded in absolute value by
$C(t)^2/\omega_n^2$, it is negligible, in particular
compared with the linear term in Eq. (\ref{car-w}).
Therefore, the functions $\bar{W}_{n}(t)$ defined in
formula (\ref{lin-sol}) can be taken as asymptotic
solutions of Eq. (\ref{car-w}) in the limit of large
$n$, up to higher order corrections.

Finally, we can show that the unitary implementability
condition is indeed fulfilled. From our arguments
above and the fact that $g_n=(n+1)^2$ and
$\omega^2_n=n(n+2)$, it is clear that the summand in
condition (\ref{newnewcondition}) has the same
asymptotic behavior as $|W_n|^2$ for large $n$. But
$|W_n|^2$ is certainly summable since, apart from
subdominant terms which do not affect the summability,
$|W_n|^2$ is bounded by $C(t)^2/\omega_n^2$. So, the
unitary implementability condition is satisfied.

\section{Uniqueness of the quantization}
\label{sec:3}

We have just seen that there is a Fock representation,
determined by an $SO(4)$-invariant complex structure,
which allows a unitary dynamics for the scalar field
with a time dependent mass. The question we want to
address now is that of uniqueness. Are there distinct
representations with the same properties? We will show
that the answer is in the negative. Although there are
in fact infinitely many distinct, i.e., non-unitarily
equivalent, $SO(4)$-invariant Fock representations, we
will show that the requirement of a unitary
implementation of the field dynamics selects in fact a
unique unitary equivalence class of representations.

\subsection{Invariant complex structures}
\label{inva}

Let us restrict our considerations from now on
exclusively to complex structures $J$ that are
invariant under the action of the symmetry group,
i.e., such that $T^{-1}JT=J$, $\forall T \in SO(4)$.
We will refer to such complex structures simply as
invariant ones.

A characterization of invariant complex structures is
easily obtained by employing Schur's lemma  as
follows (see Ref. \cite{BVV2}). The phase space
$\Gamma$ can be decomposed as the direct sum
\begin{equation}
\Gamma=\bigoplus_n \Gamma_n,
\end{equation}
with $\Gamma_n:={\cal Q}_n\oplus {\cal P}_n$. Then, a
simple application of Schur's
lemma \cite{schur} leads
to the conclusion that an invariant complex structure
$J$ is block diagonal with respect to the above
decomposition, i.e., an invariant complex structure
$J:\Gamma\to \Gamma$ is of the form
\begin{equation}
\label{jn}
J=\bigoplus_n J_n,
\end{equation}
where $J_n:\Gamma_n \to \Gamma_n$ are arbitrary
$SO(4)$-invariant complex structures. Consider now a
basis in each space $\Gamma_n$, for instance the basis
provided by the variables $q_{n\ell m}$ and the
corresponding momenta $p_{n\ell m}$. Then, each $J_n$
corresponds to a matrix, characterized by four square
blocks: $J^{qq}_n$, $J^{qp}_n$, $J^{pq}_n$, and
$J^{pp}_n$, which connect respectively ${\cal Q}_n$ to
itself, ${\cal Q}_n$ to ${\cal P}_n$, ${\cal P}_n$ to
${\cal Q}_n$, and ${\cal P}_n$ to itself. The
invariance conditions on those blocks  reads
\begin{equation}
\label{26}
D_q(T)J^{qq}_n=J^{qq}_nD_q(T), \qquad
D_p(T)J^{qp}_n=J^{qp}_nD_q(T),
\end{equation}
\begin{equation}
\label{27}
\ D_q(T)J^{pq}_n=J^{pq}_nD_p(T), \qquad
D_p(T)J^{pp}_n=J^{pp}_nD_p(T),
\end{equation}
where $T$ is an arbitrary $SO(4)$ transformation and
$D_q$ (respectively $D_p$) denotes the matrix
representation realized in ${\cal Q}_n$ (${\cal
P}_n$). Actually, {according to our comments at the
end of Sec. \ref{secfi},} the matrices $D_q(T)$ and
$D_p(T)$ must coincide for each $T$, {so that}
conditions (\ref{26}, \ref{27}) become invariance
conditions for matrices in the considered
$(n+1)^2$-dimensional representation of $SO(4)$, which
is known to be irreducible. Then, again by Schur's
lemma, each of the matrices $J^{qq}_n$, $J^{qp}_n$,
$J^{pq}_n$, and $J^{pp}_n$ must be proportional to the
identity matrix ${\bf I}$, namely $J^{qq}_n=a_n {\bf
I}$, $J^{pq}_n=b_n {\bf I}$, $J^{qp}_n=c_n {\bf I}$,
and $J^{pp}_n=d_n {\bf I}$, with $a_n$, $b_n$, $c_n$,
and $d_n$ certain complex numbers.

{In fact, the above conclusion applies in principle
to} complex representations only, whereas we are
dealing {instead} with the real spaces ${\cal Q}_n$
and ${\cal P}_n$. {Nonetheless, it} is clear that
every invariant complex structure {in these} real
spaces can be obtained from {the} restriction (to
{such} real subspaces) of an invariant complex
structure {defined on their corresponding complex
counterparts}. The above result {is therefore valid
in our case as well}. Moreover, since only real
matrices have a well defined action on real subspaces,
the proportionality coefficients $a_n$, $b_n$, $c_n$,
and $d_n$ {must be all} real. Finally, since $J$ is
a complex structure, one can conclude that, for all
$n$, these coefficients, organized in the matrix
\begin{eqnarray}
\label{ma}
\left( \begin{array}{cc} a_n
& b_n \\ c_n &
d_n \end{array} \right),
\end{eqnarray}
must form a complex structure in two dimensions.

Once we have characterized the invariant complex
structures it is convenient, for what follows, to
relate them to the complex structure $j_0$ which
determines the representation (\ref{mu}, \ref{rep}) of
Sec. \ref{repsec}. We start by decomposing $\Gamma_n$
in terms of independent degrees of freedom, i.e., as
the direct sum
\begin{equation}
\label{dec}
\Gamma_n=\bigoplus_{\ell m}\Gamma_{n\ell m},
\end{equation}
where each $\Gamma_{n\ell m}$ is the 2-dimensional
symplectic space spanned by $q_{n\ell m}$ and
$p_{n\ell m}$. It is clear from the above description
of the invariant complex structures that each $J_n$ in
Eq. (\ref{jn}) can be further decomposed in the block
diagonal form
 \begin{equation}
J_n=\bigoplus_{\ell m}J_{n\ell m},
\end{equation}
and that each transformation  $J_{n\ell m}$, for fixed
$n$, is given by the {\em same} matrix, namely the
matrix (\ref{ma}), independent of $\ell$ and $m$.

In each space $\Gamma_{n\ell m}$, we now change from
the real basis $(q_{n\ell m},p_{n\ell m})$ to the
complex variables $a_{n\ell m}$ (\ref{basic-var}) and
$a^*_{n\ell m}$, and call $j_n$ the matrix
corresponding to the $2\times 2$ complex structure
(\ref{ma}) in the new basis. It is not difficult to
show \cite{ccmv} that each $j_n$ is related to
${j_0}_n$ via a symplectic transformation ${\cal
K}_n$: $j_n={\cal K}_n{j_0}_n{\cal K}_n^{-1}$, where
${j_0}_n$ is the basic $2\times 2$ block for the
complex structure $j_0$, which in the variables
$(a_{n\ell m}, a^*_{n\ell m})$ takes the form
\begin{eqnarray}
{j_0}_n= \left( \begin{array}{cc} i & 0
\\ 0 & -i \end{array} \right).
\end{eqnarray}
In addition, we adopt the following notation for the
$2\times 2$ symplectic transformation ${\cal K}_n$:
\begin{eqnarray}
\label{kn}
{\cal K}_n= \left( \begin{array}{cc} \kappa_n &
\lambda_n
\\ \lambda_n^* & \kappa_n^* \end{array} \right),
\end{eqnarray}
with $|\kappa_n|^2-|\lambda_n|^2=1$, $\forall n \in
\mathbb{N}^+$.

In other words, any invariant complex structure $J$ is
related to $j_0$ by a correspondence of the form $J=K
j_0 K^{-1}$, where $K$ is a symplectic transformation
which is block diagonal with respect to the
decomposition (\ref{dec}), and whose basic blocks
(independent of $\ell$ and $m$) are the matrices
${\cal K}_n$ (\ref{kn}). In particular, the
information determining a given invariant complex
structure $J$ is encoded in the corresponding sequence
of matrices ${\cal K}_n$.

\subsection{Unitary dynamics and equivalence of
representations}

A given  symplectic transformation $R$ admits a
unitary implementation with respect to the complex
structure $J=K j_0 K^{-1}$ if and only if the
transformation $K^{-1}RK$ is unitarily implementable
with respect to $j_0$ \cite{ccmv}. Hence, the time
evolution, specified by the sequence of matrices
$\{{\cal{U}}_n\}$ (\ref{bogo-transf}), is unitarily
implementable with respect to the Fock representation
determined by an invariant complex structure $J$ if
and only if the $j_0$-representation admits a unitary
implementation of the symplectomorphism specified by
the sequence $\{{\cal K}_n^{-1}{\cal{U}}_n {\cal
K}_n\}$, where $\{{\cal K}_n\}$ corresponds to $J$.

It is worth noticing that, like the original evolution
matrices ${\cal{U}}_n $, the transformed ones ${\cal
K}_n^{-1}{\cal{U}}_n {\cal K}_n$ are independent of
the additional indices $\ell$ and $m$, a property
which is due precisely to the similar independence
displayed by the matrices of the symplectic
transformation $K$ {that characterizes} an
invariant complex structure. Thus, the transformed
evolution matrices have again the general form
(\ref{bogo-transf}), and a straightforward computation
allow us to relate the new matrix elements with
$\alpha_n$ and $\beta_n$, given in Eqs. (\ref{alpha},
\ref{beta}). In particular, for the new beta
coefficients, which we will call $\beta^J_n$, we
obtain:
\begin{equation} \label{beta-J}
\beta^J_n(t,t_0)=(\kappa_n^*)^2\beta_n
(t,t_0)-\lambda_n^2\beta_n^* (t,t_0)+2 i
\kappa_n^*\lambda_n {\Im}[\alpha_n(t,t_0)]\, , \qquad
\forall t\in \mathbb{I}.
\end{equation}

As expected, the unitary implementability condition
for the field dynamics, with respect to the complex
structure $J=Kj_0 K^{-1}$, amounts to the square
summability of the sequences
$\sqrt{g_{n}}\beta^J_{n}(t,t_0)$, summability which
replaces the corresponding condition
(\ref{condition}).

On the other hand, we recall that the Fock
representation specified by $J=K j_0 K^{-1}$ and the
$j_0$-representation are unitarily equivalent if and
only if the transformation $K$ can be unitarily
implemented, a requirement which reduces to the
condition
\begin{equation}
\label{lambda}
\sum_{n\ell m}|\lambda_n|^2=
\sum_{n=1}^{\infty}g_n|\lambda_n|^2<\infty,
\end{equation}
where the coefficients $\lambda_n$ are the ones
appearing in the matrices ${\cal K}_n$ given in Eq.
(\ref{kn}) (see Ref. \cite{ccmv} for more details). In
the remaining of the present section, we will show
that, if the sequences
$\{\sqrt{g_n}\beta^J_n(t,t_0)\}$  are square summable
$\forall t \in \mathbb{I}$, then the same must happen
with the sequence $\{\sqrt{g_n}\lambda_n\}$. So, we
will prove that any other representation, defined by
an invariant complex structure and allowing a unitary
implementation of the field dynamics, is in fact
unitarily equivalent to the $j_0$-Fock representation.

\subsection{Proof of uniqueness}
\label{proofuni}

According to our discussion, we restrict our
considerations to any of the invariant complex
structures $J$ such that
$\{\sqrt{g_n}\beta^J_n(t,t_0)\}$ is square summable
$\forall t\in \mathbb{I}$. Since $|\kappa_n|>1$, the
sequences
\begin{equation}
\frac{\sqrt{g_n}\beta^J_n(t,t_0)}{(\kappa_n^*)^2}
=\sqrt{g_n}\beta_n(t,t_0)-\left(\frac{\lambda_n}
{\kappa_n^*}\right)^{2}\sqrt{g_n}\beta_n^* (t,t_0)+2
i\sqrt{g_n} \left(\frac{\lambda_n}{\kappa_n^*}\right)
{\Im}[\alpha_n(t,t_0)]
\end{equation}
are then also square summable. Moreover, from the fact
that $|\kappa_n|^2-|\lambda_n|^2=1$ it follows that
the sequence $\{\lambda_n/ \kappa_n^*\}$ is bounded,
and since we already know that the sequence
$\{\sqrt{g_n}\beta_n (t,t_0)\}$ is square summable, we
can guarantee that
\begin{equation} \left\{\sqrt{g_n}\beta_n
(t,t_0)-\left(\frac{\lambda_n}{\kappa_n^*}\right)^2
\sqrt{g_n}\beta_n^* (t,t_0)\right\}\end{equation} is a
square summable sequence $\forall t\in \mathbb{I}$.
Given that the set of square summable sequences is a
linear space, one is led to the conclusion that the
sequence
\begin{equation}\label{impseq}
\left\{\sqrt{g_n}\,\frac{\lambda_n}
{\kappa_n^*}\,{\Im}[\alpha_n(t,t_0)]\right\}
\end{equation}
is square summable $\forall t\in \mathbb{I}$ as well.

In order to proceed further, we write
${\Im}[\alpha_n]$ using Eqs. (\ref{alpha}) and
(\ref{newze1}) in the form
\begin{equation}
\label{42} {\Im}[\alpha_n]=
e^{\omega_n\Re[\Theta_n]}\left(1-\frac{\Im[W_n]}
{2\omega_n}\right)\sin(\omega_n\Im[\Theta_n])+
e^{\omega_n\Re[\Theta_n]}
\frac{\Re[W_n]}{2\omega_n}\cos(\omega_n\Im[\Theta_n]).
\end{equation}
On the other hand, our asymptotic analysis about the
square summability of $\beta_n$, carried out in Sec.
\ref{asy}, ensures that both
\begin{equation}\label{factors}
\sqrt{g_n}e^{\omega_n\Re[\Theta_n]}
\frac{\Im[W_n]}{\omega_n}\quad {\rm and} \quad
\sqrt{g_n}e^{\omega_n\Re[\Theta_n]}\frac{\Re[W_n]}
{\omega_n}\end{equation} provide square summable
sequences. Returning then to the sequence
(\ref{impseq}), one concludes that the contribution
from the terms proportional to the factors in Eq.
(\ref{factors}) gives a square summable part, since
the remaining factors, namely $\lambda_n/ \kappa_n^*$
and the trigonometric functions, are bounded.
Therefore, invoking again the linearity of the space
of square summable sequences, {and using in
addition that the sequence $\{e^{\omega_n
\Re[\Theta_n]}\}$ is bounded from below because
$\omega_n\Re[\Theta_n]$ is at least of order $1/n$
according to our asymptotic analysis}, we conclude
that, for any invariant complex structure which leads
to a unitary dynamics, the sequence
\begin{equation}
\left\{\sqrt{g_n}\,\frac{\lambda_n}{\kappa_n^*}
\,\sin(\omega_n\Im[\Theta_n])\right\}\end{equation}
has to be square summable $\forall t \in \mathbb{I}$.

We next analyze the trigonometric function appearing
in this last expression. Recalling again our
asymptotic analysis of Sec. \ref{asy}, and
{provided that the second derivative of the
function $f(t)$ exists and is integrable in every
closed subinterval of $\mathbb{I}$, we can integrate
by parts the last term in Eq. (\ref{newze2}) to show
that, apart from terms of order $1/n^2$, $\Im[W_n]$ is
given by}
\begin{equation}
-\frac{f(t)}{2\omega_n}+\frac{f(t_0)}{2\omega_n}
\cos{[2\omega_n(t-t_0)]}.
\end{equation}
Since
\begin{equation} \int^t_{t_0}\d \bar{t}
\cos{[2\omega_n(\bar{t}-t_0)]}=
\frac{\sin{[2\omega_n(t-t_0)]}}{2\omega_n},\end{equation}
one can check then from Eq. (\ref{zeca}) that, except
for a remaining contribution of order $1/n^2$,
$\Im[\omega_n\Theta_n(t,t_0)]$ behaves like
\begin{equation}
\omega_n(t_0-t)+\int^{t_0}_t \d\bar{t}
\,\frac{f(\bar{t})}{2\omega_n}.\end{equation} Hence,
in the large $n$-limit we obtain
\begin{equation}
\sin(\omega_n\Im[\Theta_n]) =-
\sin{\left[\omega_n(t-t_0)+\int_{t_0}^t
\d\bar{t}\,\frac{f(\bar{t})}{2\omega_n}\right]}
+o\left(\frac{1}{n^2}\right),
\end{equation}
where $o(1/n^2)$ denotes a contribution of asymptotic
order $1/n^2$. After multiplication by
$\sqrt{g_n}\left(\lambda_n/\kappa_n^*\right)$, the
last term gives a contribution of order $1/n$ which is
square summable. Therefore, it follows that the
sequence
\begin{equation}
\label{43} \left\{\sqrt{g_n}
\,\frac{\lambda_n}{\kappa^*_n} \,\sin{\left[
\omega_n(t-t_0)+\int_{t_0}^t
\d\bar{t}\frac{f(\bar{t})}{2\omega_n}\right]}\right\}
\end{equation}
is square summable $\forall t\in \mathbb{I}$.

It is convenient to introduce now the shifted time
$T:=t-t_0$, and rewrite the elements of the sequence
(\ref{43}) in the form
\begin{equation}
\label{44} \sqrt{g_n} \,\frac{\lambda_n}{\kappa^*_n}
\,\sin{\left[ \omega_nT+\int_{0}^T
\d\bar{t}\frac{f(\bar{t}+t_0)}{2\omega_n}\right]}.
\end{equation}
Then, the function
\begin{equation}
\label{45} z(T):=\lim_{M\to
\infty}\sum_{n=1}^{M}g_n\frac{|\lambda_n|^{2}}
{|\kappa_n|^{2}}\sin^{2}{\left[ \omega_nT+\int_{0}^T
\d\bar{t}\frac{f(\bar{t}+t_0)}{2\omega_n}\right]}
\end{equation}
exists for all $T$ in the domain $\mathbb{\bar{I}}$,
obtained from the domain $\mathbb{I}$ of $t$ after the
shift by $t_0$. In particular, $z(T)$ is well defined
on some closed subinterval of the form
$\mathbb{\bar{I}}_L=[a,a+L]\subseteq \mathbb{\bar{I}}$
(for a suitable choice of the time $a$), where $L$ is
some finite positive number strictly smaller than the
length of $\mathbb{I}$.

Let us now apply Luzin's theorem \cite{luzin} to the
function $z(T)$. This theorem guarantees that, for
every $\delta>0$, there exist: i) a measurable set
$E_{\delta} \subset \mathbb{\bar{I}}_L$ such that its
complement $\bar{E}_{\delta}$ with respect to
$\mathbb{\bar{I}}_L$ satisfies
{$\int_{\bar{E}_{\delta}}\d T<\delta$}, and ii) a
function $F_{\delta}(T)$ which is continuous on
$\mathbb{\bar{I}}_L$ and coincides with $z(T)$ in
$E_{\delta}$. In particular, defining
$I_{\delta}:=\int_{E_{\delta}}F_{\delta}(T)\d T$,
which is a finite real number, we obtain from Luzin's
theorem that
\begin{equation}
\label{46}
\sum_{n=1}^{M}g_n\frac{|\lambda_n|^{2}}{|\kappa_n|^{2}}
\int_{E_{\delta}}\sin^{2}{\left[ \omega_nT+\int_{0}^T
\d\bar{t}\frac{f(\bar{t}+t_0)}{2\omega_n}\right]}\d
T\leq \int_{E_{\delta}}z(T)\d T=I_{\delta}, \qquad
\forall M \in \mathbb{N}^+ .
\end{equation}

In the appendix we prove that, actually, the above
inequality allows one to demonstrate that the sequence
$\{\sqrt{g_n} \lambda_n/\kappa_n\}$ is square
summable. From this summability and the relation
$|\kappa_n|^2-|\lambda_n|^2=1$, one can also see that
the sequence $\{\kappa_n\}$ tends to 1, and thus it is
bounded. Therefore, the sequence $\{\sqrt{g_n}
\lambda_n\}$ must be square summable, so that the
condition of unitary equivalence between complex
structures [see Eq. (\ref{lambda})] is indeed
satisfied. This concludes the proof of uniqueness of
the equivalence class of invariant complex structures
which provide a unitary implementation of the
dynamics.

\section{Conclusions and extension to other manifolds:
the case of $S^2$}
\label{sec:conc-dis}

We have considered a free scalar field with a time
dependent mass propagating in a $(1+3)$-spacetime with
the topology of $\mathbb{I}\times S^3$, where
$\mathbb{I}$ is an interval of the real line, provided
with a product metric given by the standard metric on
the three-sphere and the trivial metric on the time
interval $\mathbb{I}$. Among the infinitely many
possible Fock representations of the CCR's for this
field, we have succeded to show that the
representation which is naturally associated with the
massless free case does not only incorporate the
symmetries of the field equation, but implements the
dynamics corresponding to the time varying mass as
unitary transformations as well. More importantly, we
have proved that, among all the Fock representations
based on complex structures that are invariant under
the group of symmetries of the field equation, the
unitary equivalence class of the representation for
the free massless case is in fact the unique class
which supports a unitary dynamics. In this sense, the
Fock quantization is unique once one demands a natural
implementation of the symmetries of the field equation
and that the dynamics be unitary.

The discussion that we have presented can be extended
or generalized to cases in which the spatial manifold
is not $S^3$ but {another} compact manifold. Let us
assume that the negative of the Laplace-Beltrami
operator for that manifold (a nonnegative operator)
has also a discrete spectrum, with eigenvalues
$\omega_n^2$ labeled in increasing order by the index
$n=0,1,\ldots$, and such that $\omega_n$ tends to
infinity when $n$ does so. Again, we call $g_n$ the
dimension of each of the corresponding eigenspaces. It
is not difficult to realize that the asymptotic
analysis for large $n$ carried out in Sec. \ref{asy}
can be alternatively understood as an asymptotic
analysis for large $\omega_n$, with asymptotic
expansions in powers of $1/\omega_n$. Then, from the
arguments presented above (and recalling in particular
that $W_n$ is of order $1/\omega_n$ up to subdominant
terms), one concludes that the complex structure of
the free massless case leads to a Fock representation
with a unitary implementation of the dynamics also in
the presence of a time dependent mass (at least for a
sufficiently regular time dependence) if and only if
the sequence with elements $g_n/\omega_n^4$ is
summable. On the other hand, as we have commented, the
Fock representation associated with the free massless
case clearly provides a natural unitary implementation
of the symmetries of the Laplace-Beltrami operator.

Summarizing, the availability of the Fock
representation of the free massless case in order to
obtain an invariant vacuum and a quantum unitary
dynamics when the mass varies in time is maintained
for more general compact manifolds than $S^3$, and the
relevant condition is the summability of the sequence
formed by $g_n/\omega_n^4$. Concerning the uniqueness
of this representation, modulo unitary equivalence, we
note that the only additional property employed in our
discussion of Sec. \ref{inva}, and which has proved to
be {\em sufficient} to guarantee uniqueness, is the
irreducibility of the representation of the group of
symmetries of the field equation in each of the
eigenspaces of the Laplace-Beltrami operator. It is
worth emphasizing that this condition suffices to
reach uniqueness once the Fock representation which
corresponds to the free massless case has been seen to
be admissible (i.e., provides a unitary dynamics), but
the condition itself is {\em not necessary}.

We are now in an adequate position to prove that our
results about the unitary implementation of the
dynamics and the uniqueness of the equivalence class
of the representation associated with the massless
situation are valid as well for the case of the
two-sphere, as we had anticipated. If the spatial
manifold is $S^2$, equipped with the standard metric,
all our assumptions are fulfilled, with eigenvalues
given by $\omega_n^2= n(n+1)$, degeneracy factors
equal to $g_n=2n+1$, and symmetry group given by
$SO(3)$. Note in particular that $g_n/\omega_n^4$
behaves asymptotically as $1/n^3$, so that the
corresponding sequence is indeed summable. Therefore,
unitarity and uniqueness of the considered
representation are reached for the two-sphere without
appealing to the axisymmetry of the field,
generalizing in this way the conclusions of Ref.
\cite{cmsv} where the axisymmetric restriction was
studied.

It is also worth commenting that the summability
condition for $g_n/\omega_n^4$ appears to depend
importantly on the dimensionality. Generically, for a
compact spatial manifold of dimension $d$, one would
expect that the dimension $g_n$ of the eigenspaces of
the Laplace-Beltrami operator increases for large $n$
as $n^{d-1}$, {except if there are suppressions in
this degeneracy owing to the specific symmetry of the
system. Similarly, one would expect that} the
eigenvalue $\omega_n^2$ {increases} quadratically,
as $n^2$. {All this} is actually so for the
two-sphere and the three-sphere. Then, the requirement
that $g_n/\omega_n^4$ be summable would amount to the
summability of the sequence of elements $n^{d-5}$,
which is true only in (integer) dimensions equal or
smaller than three. These {heuristic} arguments
seem to indicate that, in addition to the compactness
of the spatial manifold, the dimensionality plays a
fundamental role for the unitary implementation of the
dynamics in the Fock representation obtained from the
massless field case.

One interesting situation, owing to its application to
cosmological perturbations, like in the case of the
three-sphere, is when the spatial manifold is the
three-torus (equipped with the Euclidean metric). The
corresponding eigenvalues of the Laplace-Beltrami
operator are labeled by three integers ($n_1$, $n_2$,
and $n_3$), one for each dimension, and are given by
$\omega_n^2=n_1^2+n_2^2+n_3^2$. It is possible to
check then the summability of the sequence
corresponding to elements of the form
$g_n/\omega_n^4$. On the other hand, the group of
symmetries of the field equation is the product of the
three copies of the group of translations in the
circle. Note that the representation of this group is
not irreducible in the eigenspaces of the
Laplace-Beltrami operator. Nonetheless, as we have
remarked, this requirement of irreducibility is not a
necessary condition to reach the uniqueness of the
representation. Given that the representation is
irreducible only in the sectors determined by the
whole set of indices $(n_1,n_2,n_3)$, the
symplectomorphism $K$ that characterizes the invariant
complex structures will now decompose in matricial
blocks labeled by $(n_1,n_2,n_3)$, rather than by $n$.
The corresponding $2\times 2$ matrices ${\cal K}_{n_1
n_2 n_3}$ will now cease to be the same for all the
blocks associated with the same eigenspace of the
Laplace-Beltrami operator (i.e., they will not depend
only on $n$). In spite of this, it is possible to show
that the proof of Sec. \ref{sec:3} can be generalized
taking into account this fact and conclude again the
uniqueness of the Fock representation. We will address
this issue in more detail elsewhere.

Finally, let us remark that the proof of uniqueness
that we have presented allows one to select a
preferred Fock quantization in the list of scenarios
and situations discussed in the Introduction of this
article. In particular, it is worth noting that the
kind of rescaling of the cosmological perturbations by
an appropriate time function (determined by the FRW
scale factor) which leads to a field equation of the
type considered in this work has been frequently
employed in the literature to pass to the quantum
description of these perturbations (see e.g.
\cite{muk2}, although the discussion there is
specialized to the case of a flat FRW background). At
least for compact spatial topology, the results of our
work demonstrate that the Fock quantization adopted in
those {cases} is in fact the (essentially) unique
one which incorporates in a natural way the symmetries
of the field equation (as symmetries of the vacuum)
and provides a unitary implementation of the quantum
dynamics for a fixed classical background.

\section*{Acknowledgements}
{G.A.M.M. is grateful to Javier Olmedo for
suggestions and fruitful discussions.} This work was
supported by the Spanish MICINN Project
FIS2008-06078-C03-03, the Spanish Consolider-Ingenio
2010 Program CPAN (CSD2007-00042), {and the DGAPA-UNAM
(Mexico) Grant No. IN108309-3.}

\appendix
\section{Calculations for the proof of uniqueness}

In this appendix we will show that inequality
(\ref{46}) suffices to prove the square summability of
the sequence $\{\sqrt{g_n} \lambda_n/\kappa_n\}$, a
fact from which one can deduce the unitary equivalence
of all the invariant complex structures which
implement the dynamics as a unitary transformation.

We will first show that Eq. (\ref{46}) provides us
with a bound on $\sum g_n
|\lambda_n|^{2}/|\kappa_n|^{2}$. For that, we note
that
\begin{eqnarray}
\label{47} \nonumber \int_{E_{\delta}}\sin^{2}{\left[
\omega_nT+\int_{0}^T
\d\bar{t}\frac{f(\bar{t}+t_0)}{2\omega_n}\right]}\d
T \\
\nonumber =\int_{\mathbb{\bar{I}}_L} \sin^{2}{\left[
\omega_nT+\int_{0}^T
\d\bar{t}\frac{f(\bar{t}+t_0)}{2\omega_n}\right]} \d
T-\int_{\bar{E}_{\delta}}\sin^{2}{\left[
\omega_nT+\int_{0}^T
\d\bar{t}\frac{f(\bar{t}+t_0)}{2\omega_n}\right]} \d T
\\ \geq\int_{\mathbb{\bar{I}}_L}\sin^{2}{\left[
\omega_nT+\int_{0}^T
\d\bar{t}\frac{f(\bar{t}+t_0)}{2\omega_n}\right]} \d T
-\delta, \qquad \forall n.
\end{eqnarray}
We recall that $E_{\delta}$ is the set introduced in
the discussion of Luzin's theorem, while
$\bar{E}_{\delta}$ is its complement in
$\mathbb{\bar{I}}_L$, its measure being smaller than
$\delta$. In addition, for the integral over
${\mathbb{\bar{I}}_L}$ we find
\begin{eqnarray}
\label{48} \nonumber \int_{\mathbb{\bar{I}}_L}
\sin^2\left[ \omega_nT+\int_{0}^T \d\bar{t}
\frac{f(\bar{t}+t_0)}{2\omega_n}\right]\d T=
\frac{L}{2} - \frac{\sin{\left[
2\omega_n(a+L)+\int_{0}^{(a+L)} \d\bar{t}
\frac{f(\bar{t}+t_0)}{\omega_n}\right]}}{4\omega_n+
2f(a+L+t_0)/\omega_n} \\
\nonumber
 + \frac{\sin{\left[2 \omega_na+\int_{0}^{a} \d\bar{t}
\frac{f(\bar{t}+t_0)}{\omega_n}\right]}}
{4\omega_n+2f(a+t_0)/\omega_n} -
\frac{1}{8\omega_n^3}\int_{\mathbb{\bar{I}}_L}
\frac{f^{\prime}(T+t_0)}{\left[1+\frac{f(T+t_0)}
{2\omega_n^2}\right]^2}
\sin{\left[2\omega_nT+\int_0^T \d\bar{t}
\frac{f(\bar{t}+t_0)}{\omega_n}\right]} \d T \\
\nonumber \geq\frac{L}{2}-\frac{\omega_n}{|4\omega_n^2
+2f(a+L+t_0)|}-\frac{\omega_n}{|4\omega_n^2+2f(a+t_0)|}
-\frac{1}{8\omega_n^3}\int_{\mathbb{\bar{I}}_L}
\frac{|f^{\prime}(T+t_0)|}{\left[1+\frac{f(T+t_0)}
{2\omega_n^2}\right]^2}\d T.
\end{eqnarray}
In the following, we restrict ourselves to $n\geq
n_0$, where $n_0$ is any fixed (positive) integer such
that $\omega_{n_0}^2$ is larger than the maximum of
the function $|f(T+t_0)|/(2D)$ in the interval
${\mathbb{\bar{I}}_L}$, and $D<1$ is any fixed
constant. Then, given that $\omega_n$ increases
monotonically with $n$,
\begin{equation}
 \left|1+\frac{f(T+t_0)}{2\omega_n^2}\right| \geq
1-\frac{|f(T+t_0)|}{2\omega_n^2} \geq
1-\frac{|f(T+t_0)|}{2\omega_{n_0}^2} \geq 1-D,
\end{equation} for all $T$ in
the interval of integration, including its end points.
Hence one obtains that, for all $n\geq n_0$,
\[ \int_{\mathbb{\bar{I}}_L}  \sin^2\left[
\omega_n T+\int_{0}^T \d\bar{t}
\frac{f(\bar{t}+t_0)}{2\omega_n}\right]\d T \geq
\frac{L}{2} - \frac{1}{2 \omega_{n_0} (1-D)} -\frac{
\int_{\mathbb{\bar{I}}_L}  |f^{\prime}(T+t_0)|\d T}{8
\omega_{n_0}^3 (1-D)^2}:= \Lambda_{n_0},\] where we
also assume that $n_0$ is such that $\Lambda_{n_0}>0$
(this can always be fulfilled with an appropriate
choice, since $\Lambda_{n_0}$ tends to $L/2$ when
$n_0$ tends to infinity).

We now introduce the above result in the last
inequality of Eq. (\ref{47}). In this way, we get
\begin{equation}
\int_{E_{\delta}}\sin^{2}{\left[ \omega_nT+\int_{0}^T
\d\bar{t}\frac{f(\bar{t}+t_0)}{2\omega_n}\right]}\d T
\geq \Lambda_{n_0}-\delta.
\end{equation}
We choose $\delta$ such that $\Lambda_{n_0}>\delta$,
something which is always possible. Then, it follows
from Eq. (\ref{46}) that, for all $M\geq n_0$,
\begin{equation}
\sum_{n=n_0}^M g_n
\frac{|\lambda_n|^2}{|\kappa_n|^2} \leq \frac{
I_{\delta}}{\Lambda_{n_0}-\delta}.
\end{equation}
Since $n_0$ is fixed and the above bound is valid for
arbitrary large $M$, it follows that the sequence of
nonnegative elements $g_n\,|\lambda_n|^2/|\kappa_n|^2$
is summable, as we wanted to prove.

\end{document}